# Electrical Tuning of Valley Magnetic Moment via Symmetry Control

Sanfeng Wu[1], Jason S Ross[2], Grant Aivazian[1], Aaron Jones[1], Zaiyao Fei[1], Gui-Bin Liu[3], Wenguang Zhu[4-6], Di Xiao[4,7], Wang Yao[3], David Cobden[1], Xiaodong Xu[1,2]

[1] Department of Physics, University of Washington, Seattle, Washington 98195, USA

[2] Department of Material Science and Engineering, University of Washington, Seattle, Washington 98195, USA

[3] Department of Physics and Center of Theoretical and Computational Physics, The University of Hong Kong, Hong Kong, China

[4] Materials Science and Technology Division, Oak Ridge National Laboratory, Oak Ridge, Tennessee, 37831, USA

[5] Department of Physics and Astronomy, University of Tennessee, Knoxville, Tennessee 37996, USA

[6] ICQD/HFNL, University of Science and Technology of China, Hefei, Anhui, 230026, China

[7] Department of Physics, Carnegie Mellon University, Pittsburg, PA 15213, USA

**Crystal symmetry governs the nature of electronic Bloch states. For example, in the presence of time reversal symmetry, the orbital magnetic moment and Berry curvature of the Bloch states must vanish unless inversion symmetry is broken[1]. In certain 2D electron systems such as bilayer graphene, the intrinsic inversion symmetry can be broken simply by applying a perpendicular electric field[2-3]. In principle, this offers the remarkable possibility of switching on/off and continuously tuning the magnetic moment and Berry curvature near the Dirac valleys by reversible electrical control[4-5]. Here we demonstrate this principle for the first time using bilayer $MoS_2$, which has the same symmetry as bilayer graphene but has a bandgap in the visible[6-7] that allows direct optical probing of these Berry-phase related properties [5,8-12]. We show that the optical circular dichroism, which reflects the orbital magnetic moment in the valleys[1,8], can be continuously tuned from -15% to 15% as a function of gate voltage in bilayer $MoS_2$ field-effect transistors. In contrast, the dichroism is gate-independent in monolayer $MoS_2$, which is structurally non-centrosymmetric. Our work demonstrates the ability to continuously vary orbital magnetic moments between positive and negative values via symmetry control. This represents a new approach to manipulating Berry-phase effects for applications in quantum electronics associated with 2D electronic materials.**

**Text**

The Dirac valley degree of freedom has recently been considered for new modes of electronic and photonic device operation[4-5,9-17] due to the arrival of atomically thin two-dimensional (2D) electronic systems (Fig. 1a)[6-7,18-19]. In this context, phenomena such as valley polarization and anomalous valley- and spin-Hall effects have been discussed for the "+K" and "-K" Dirac valleys at opposite corners of the Brillouin zone in hexagonal systems (Fig. 1a)[9-12,15]. The realization of these effects hinges on achieving control of valley contrast, i.e., of properties that differ between the two valleys, in particular the magnetic moment (**m**) and Berry curvature ($\mathbf{\Omega}$), where **m** gives rise to circular dichroism[5, 8] and $\mathbf{\Omega}$ is responsible for topological transport[1]. As a result, charge carriers in the two valleys can be distinguished by their different response to optical and electric fields[9-11], providing the basis for valley-dependent operations.

Time reversal symmetry dictates that each of these pseudovectors **m** and $\Omega$ have the same magnitude but opposite sign in the two valleys, while inversion symmetry requires them to have the same sign. Thus a necessary condition for valley-contrasting **m** and $\Omega$ is inversion symmetry breaking[4]. This constraint could be revealed explicitly in a system where the inversion symmetry can be adjusted at will. However, continuous and reversible control of these valley-contrasting Berry-phase properties has not been achieved previously. In this Letter, we report the successful electrical control of valley magnetic moment between finite negative and positive values, by applying a perpendicular electric field to bilayer $MoS_2$.

Unlike monolayer $MoS_2$ with its structural inversion asymmetry[9,20-21], pristine bilayer $MoS_2$ is inversion symmetric (Fig. 1b)[9,11-12]. Thus **m** and the associated optical dichroism all vanish in the ±K valleys. However, the inversion symmetry can be broken by an external perturbation such as an electric field applied perpendicular to the bilayer, which leads to a potential difference between the two layers (row 3 in Fig. 1b). In bilayer graphene this electric field determines the bandgap[2-3]. In bilayer $MoS_2$, which already has a bandgap in the visible regime[6-7,19], the effect on the energy spectrum is negligible. This permits us focus on the effect of electric field on the orbital magnetic moments, which can be optically probed using the valley-contrasting circular dichroism near the ±K valleys (Fig. 1c)[9]. In our discussion below, all Berry-phase related physical quantities should be taken to be valley-contrasting.

In our experiments, atomically thin $MoS_2$ samples are mechanically exfoliated from bulk $MoS_2$ crystals onto $SiO_2$ (285 nm thick) on heavily doped silicon[18]. Figure 1d is an optical micrograph of a representative sample. The number of layers is identified by atomic force microscopy, as illustrated in Figs. 1e and f, where the profile along the white dashed line clearly shows monolayer steps with thickness[7,17] about 7 Å. We then fabricate field-effect transistors (FETs) in which the back-gate voltage $V_g$ controls the perpendicular electric field (Fig. 1g)[17,22-24] (see Methods).

We use polarization-resolved micro-photoluminescence (PL) to probe the relevant magnetoelectric effect, i.e., the electric-field induced orbital magnetic moment. To explain the technique we consider the case of right circularly polarized ($\sigma^+$) excitation (see Methods for details). The excitation laser is incident normally on the sample, held at a temperature of 30 K, with a spot size of ~ 2 μm. The laser intensity is 150 W/cm$^2$ and the wavelength is 632 nm unless otherwise noted. After valley polarization is generated through interband excitation by $\sigma^+$ light, the right- and left-handed polarized PL signals, $P(\sigma^+)$ and $P(\sigma^-)$, are selectively detected. The degree of circular polarization, $\eta = \frac{P(\sigma^+)-P(\sigma^-)}{P(\sigma^+)+P(\sigma^-)}$, gives a measure of circular dichroism and thus reflects the magnitude of the orbital magnetic moments generated by the controlled electric field. All PL spectra presented here show a sharp drop below 645 nm due to the spectral cut-off of the laser notch filter.

We start with the key result of this Letter: electrical control of the orbital magnetic moment in bilayer $MoS_2$, manifested as a continuous tuning of the circular dichroism by gate voltage $V_g$. Figures 2a and b show the polarization-resolved PL spectra on excitation by $\sigma^+$ and $\sigma^-$ polarized light, respectively, at $V_g = 0$ for device B1. There are two spectral features. The broad peak centered on 860 nm results from phonon-assisted indirect interband transitions, which should not be circularly polarized, as is the case. This peak is not the focus of the paper but it provides a convenient non-polarized reference signal. In contrast, the PL from the ±K valleys (at 650 nm) does show circular polarization. For $\sigma^+$ excitation $P(\sigma^+)$ is larger than $P(\sigma^-)$ (Fig. 2a), and vice versa (Fig. 2b). The corrected polarization $\eta$ is plotted against wavelength in Fig. 2c (see methods). The increase in the magnitude of $\eta$ towards shorter wavelength is characteristic of hot luminescence. The nonzero PL polarization at $V_g = 0$ implies the existence of broken inversion symmetry in the as-prepared bilayer, consistent with a recent report[11]. All measured bilayer devices showed $\eta$ ranging from ~10% to ~30% at $V_g = 0$ (Supplementary Materials), with no clear dependence on laser intensity.

Figure 2d shows the 2D intensity plot of $P(\sigma^+)$ on $\sigma^+$ excitation for device B2. The light emission is slightly blue-shifted and its intensity decreases as $V_g$ decreases from positive to negative (Supplementary Materials). The sharp spikes superimposed on the PL are Raman scattering peaks[25-27]. The remarkable observation is that the polarization of PL from the bilayer changes drastically as a function of $V_g$. Figure 2e is a color map of $\eta$ as a function of both wavelength and $V_g$. The data clearly show that $\eta$ depends strongly on $V_g$ and approaches zero around -60 V. This observation implies that the sample is initially electron doped[17], which indicates trapped positive charge between the $MoS_2$ and substrate (Supplementary Materials).

We extract $\eta$ at 648 nm in Fig. 2e (indicated by the dashed lines) and plot it in Fig. 2f. It shows a striking "X" pattern. The red and blue curves represent $\eta$ for $\sigma^+$ and $\sigma^-$ polarized light excitation respectively. The PL polarization completely disappears at the center of the "X" at $V_c$ = -60 V, implying the restoration of inversion symmetry at this point. $\eta$ can be made finite by either increasing or decreasing $V_g$. Also, the dependence on $V_g$ is nonlinear: it appears to be approaching saturation at about 10% for $|V_g - V_c| \gtrsim 30$ V.

The X pattern dramatically demonstrates the relation between inversion symmetry and the orbital magnetic moment, which can be switched on and continuously varied between positive and negative values by gate electric field. When inversion symmetry is present in the bilayer at $V_g = V_c$, **m** and the associated optical dichroism disappear as expected. For all other values of $V_g$, as inversion symmetry is broken, valley circular dichroism is observed as the signature of nonzero **m**. This shows that in addition to modifying the energy spectrum, which has already been demonstrated in biased bilayer graphene, inversion symmetry breaking can also be used to control the Berry-phase related properties of Bloch states. The bilayer configurations at the two gate voltages $V_g = V_c \pm \Delta V$ are equivalent by the operation of spatial inversion, under which the +K and –K valley indexes are switched but the resulting PL polarization is unchanged. Thus $\eta$ is always positive (negative) for $\sigma+$ ($\sigma-$) excitation and is symmetric with respect to $V_c$, as demonstrated by the red (blue) data points in Fig. 2f. This shows that contrasting and measurable physical properties such as magnetic moment and circular dichroism provide better intrinsic labels for the valleys compared with momentum space positions which only have relative meaning and depend on crystal orientation.

In contrast to bilayer $MoS_2$, we observe no magnetoelectric effect in monolayer $MoS_2$. Five monolayer devices all showed the same behavior. Figure 3a illustrates PL spectra from device S1 on excitation by $\sigma^+$ polarized light at $V_g = 0$. The broad PL emission centered at 705 nm is probably from impurities. It does not show appreciable polarization and its intensity strongly depends on $V_g$ (see Supplementary Materials). The narrower peak at 650 nm is the emission from the ±K valleys and is strongly polarized[10-12]. The degree of polarization $\eta$ reaches as much as 0.8 at 648 nm, with some variation between devices. Such a large degree of polarization indicates strongly suppressed inter-valley scattering[10-11].

Figure 3b shows a PL intensity map for device S2. As in bilayer $MoS_2$, the sharp vertical lines here are Raman scattering peaks which are superimposed on the PL of interest. The PL intensity has an appreciable dependence on $V_g$. However, $\eta$ remains constant as $V_g$ varies from -50 V to 50 V. This is evident in the maps of $\eta$ vs $V_g$ and wavelength shown in Fig. 3c. Fig. 3d shows $\eta$ vs $V_g$ at a specific wavelength of 648 nm for the $\sigma^+$ (red) and $\sigma^-$ (blue) excitation. The lack of dependence on $V_g$ is consistent with the fact that in monolayer $MoS_2$ the valley magnetic moment is generated by the structural inversion asymmetry which is not affected by electric field.

The explicit relation between circular dichroism and orbital magnetic moment has been well established, as discussed in e.g. Ref. 5 and 8. For direct optical transitions at +K and -K valleys between a pair of bands $n$ and $s$, the momentum-resolved degree of circular polarization is given by $\eta(k) = \frac{|P_+^{ns}(k)|^2 - |P_-^{ns}(k)|^2}{|P_+^{ns}(k)|^2 + |P_-^{ns}(k)|^2}$, where $P_\pm^{ns}(k) = \langle u_s(k)|(P_x \pm iP_y)|u_n(k)\rangle$ is the interband matrix element of the momentum operator. The orbital magnetic moment in a 2D system can be expressed in terms of the Bloch functions $|u_n(k)\rangle$ as[5] $\mathbf{m_s}(k) = \hat{\mathbf{z}} \frac{\mu_B}{2m_e} \sum_{n \neq s} \frac{|P_+^{ns}(k)|^2 - |P_-^{ns}(k)|^2}{\varepsilon_n(k) - \varepsilon_s(k)}$, where $\mu_B$ is the Bohr magneton, $m_e$ is free electron mass, $\varepsilon_n(k)$ is the dispersion of the $n$'th band, and $\hat{\mathbf{z}}$ is perpendicular to the sheet. Therefore, a nonzero $\eta$ signifies the appearance of finite orbital magnetic moment at the valleys.

We have performed *ab initio* density functional theory (DFT) calculations of both $\eta$ and **m** as a function of electric field in bilayer. Figure 4a shows the calculated $\eta$ when the photo-excitation energy is ~50 meV above the DFT bandgap, which is roughly the energy difference between the incident photons and the PL peak position. The calculation qualitatively agrees with the experimental observation. Fig. 4b-c shows the calculated orbital magnetic moment at +K and -K valleys of the top valence band under various electric fields, a manifestation of the valley-contrasting magnetoelectric effect. The nonzero orbital moment also implies the appearance of finite Berry curvature[4-5,9-10], which also depends on the inter-band matrix element of the momentum operator and has a similar expression: $\mathbf{\Omega}_s(k) = -\hat{z}\frac{1}{2}\left(\frac{\hbar}{m_e}\right)^2 \sum_{n \neq s} \frac{|P_+^{ns}(k)|^2 - |P_-^{ns}(k)|^2}{(\varepsilon_n(k) - \varepsilon_s(k))^2}$. Our demonstration of the valley-contrasting magnetoelectric effect therefore provides evidence of electrical control of Berry curvature as well, pointing to a novel way of manipulating topological quantum phenomena in atomically thin 2D materials.

**Methods**

**Device Fabrication**: Devices were fabricated using standard electron beam lithography techniques, using an FEI Sirion SEM with a Nabity Nanometer Pattern Generation system. An electron beam evaporator was then used to deposit 5nm/50nm of Ti/Au followed by a standard lift-off process.

**Polarization resolved Photoluminescence**: The polarization resolved PL setup is equipped with a high power microscope (Olympus), low temperature micro-PL cryostat (Janis), and a spectrometer with a charge coupled device detector. Figure 1g shows the measurement scheme. A linearly polarized laser beam, either horizontal (H) or vertical (V), is reflected by a dichroic beam splitter (DC) and then passed through a quarter wave plate (QWP). The circularly polarized light, either $\sigma+$ or $\sigma-$, is focused by a 40X objective lens onto the sample located in the cryostat. The PL signal is collected by the same lens and goes through the same QWP. The $\sigma+$ ($\sigma-$) polarized PL is consequently converted to linear polarization V (H). The linearly polarized PL transmits through the DC and is selectively detected with a linear polarizer. The measured PL polarization is corrected due to the depolarization effects caused by the system, which mainly originate from the optics involved in the incident path and are determined by characterizing the beam polarization at the sample position. We also applied linearly polarized light to excite the device and confirmed that the resulting degree of polarization is zero.

**Acknowledgments:**

This work is mainly supported by the US DoE, BES, Division of Materials Sciences and Engineering (DE-SC0008145), and partially supported by NSF (DMR-1150719). A. Jones was supported by the NSF Graduate Research Fellowship (DGE-0718124). GL and WY were supported by Research Grant Council of Hong Kong. WZ and DX were supported by US DoE, BES, Division of Materials Sciences and Engineering. DHC and ZF were supported by the DoE BES. Device fabrication was performed at the University of Washington Micro Fabrication Facility and NSF-funded Nanotech User Facility.

**Author Contribution**:

All authors discussed the results and made critical contributions to the work.

**Figures:**

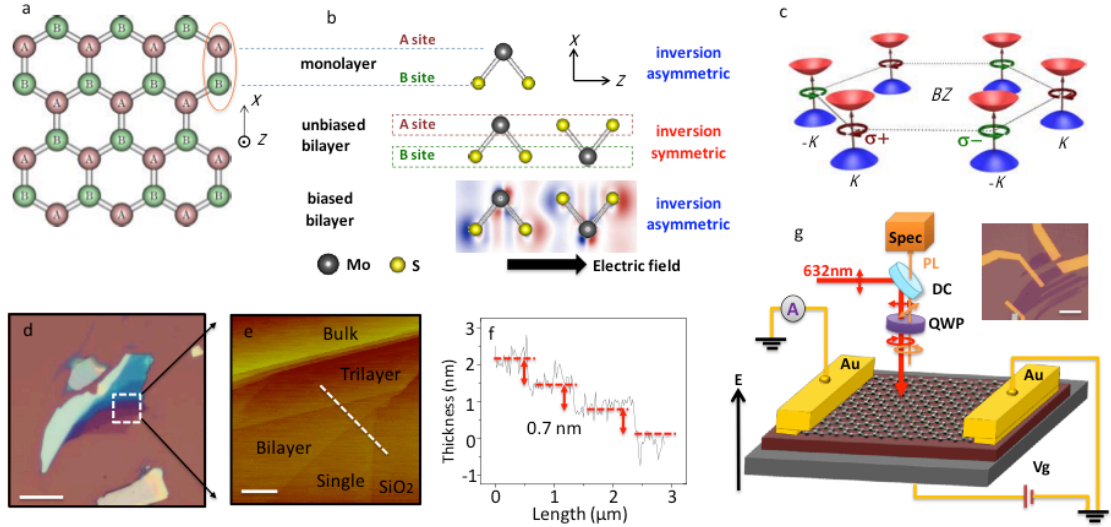

**Figure 1| MoS$_2$ Devices and their symmetry properties. a,** Top view of 2D hexagonal lattice structure. **b**, Side view of unit cells of the MoS$_2$ lattice structures. Top row: monolayer with structural inversion asymmetry. Middle row: pristine bilayer with inversion symmetry. Bottom row: change in electron density on application of a perpendicular electric field applied to the bilayer (DFT calculation), showing the absence of inversion symmetry. **c**, Cartoon of valley contrasting circular dichroism when inversion symmetry is broken. **d**, Optical microscope image of a representative MoS$_2$ sample. Scale bar: 10 μm. **e**, Atomic force microscope image of the area highlighted by the dashed square in (d). Scale bar: 1 μm. **f**, Line cut along the white dashed line in (e) shows the atomic layer thickness, ~0.7 nm. **g**, Schematic of polarized photoluminescence measurements on a MoS$_2$ FET device (see Methods). Inset: Optical microscope image of MoS$_2$ FET devices. Scale bar: 10 μm.

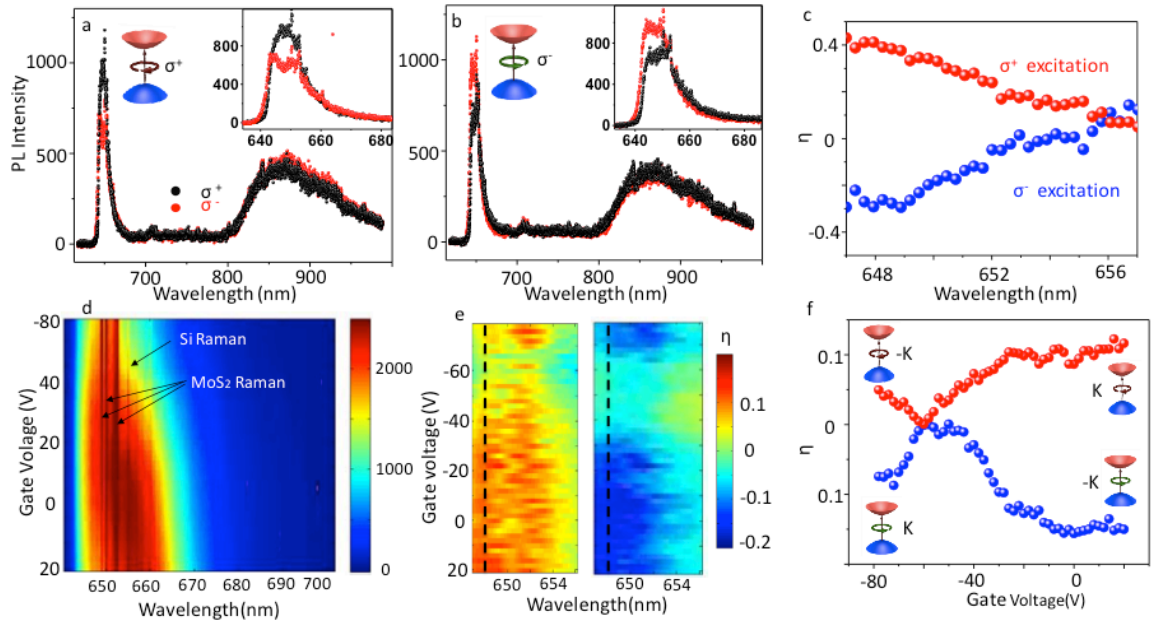

**Figure 2| Electrical control of valley magnetic moment in bilayer MoS$_2$ FETs.** Polarization resolved PL excited by **a**, $\sigma^+$ and **b**, $\sigma^-$ light at $V_g = 0$. Insets: zoomed in PL spectra around 650 nm. **c**, Degree of PL polarization as a function of wavelength. Red (blue): $\sigma^+$ ($\sigma^-$) excitation. **d**, PL intensity map as a function of wavelength and gate voltage. **e**, Degree of PL polarization as a function of wavelength and gate voltage. Left (right) plot is obtained for $\sigma^+$ ($\sigma^-$) excitation. **f**, Degree of PL polarization as function of gate voltage at 648 nm (line cuts along the dashed

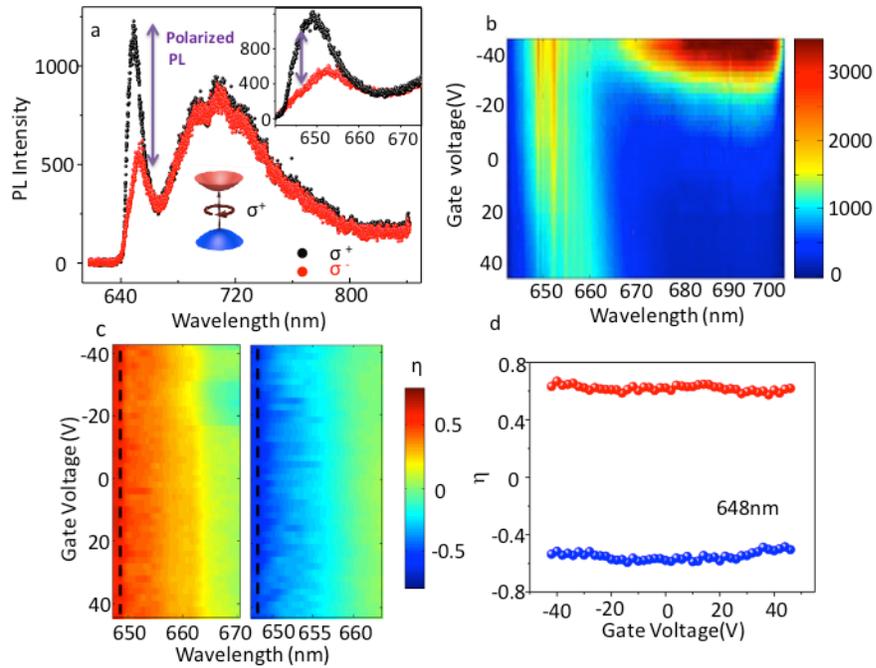

**Figure 3| Gate independent PL polarization of monolayer MoS$_2$ FETs**. **a**, Polarized PL at $V_g$=0 with σ$^+$ excitation. Black (red): σ$^+$ (σ$^-$) detection. Insets: zoom on PL spectrum near 650 nm. **b**, PL intensity map as a function of wavelength and gate voltage. **c**, Degree of PL polarization as a function of wavelength and gate voltage. Left (right) plot is obtained for σ$^+$ (σ$^-$) excitation. **d**, Degree of PL polarization at 648 nm as a function of gate voltage (indicated by the dashed lines in c). Red (blue) dots represent PL polarization generated by σ$^+$ (σ$^-$) polarized light.

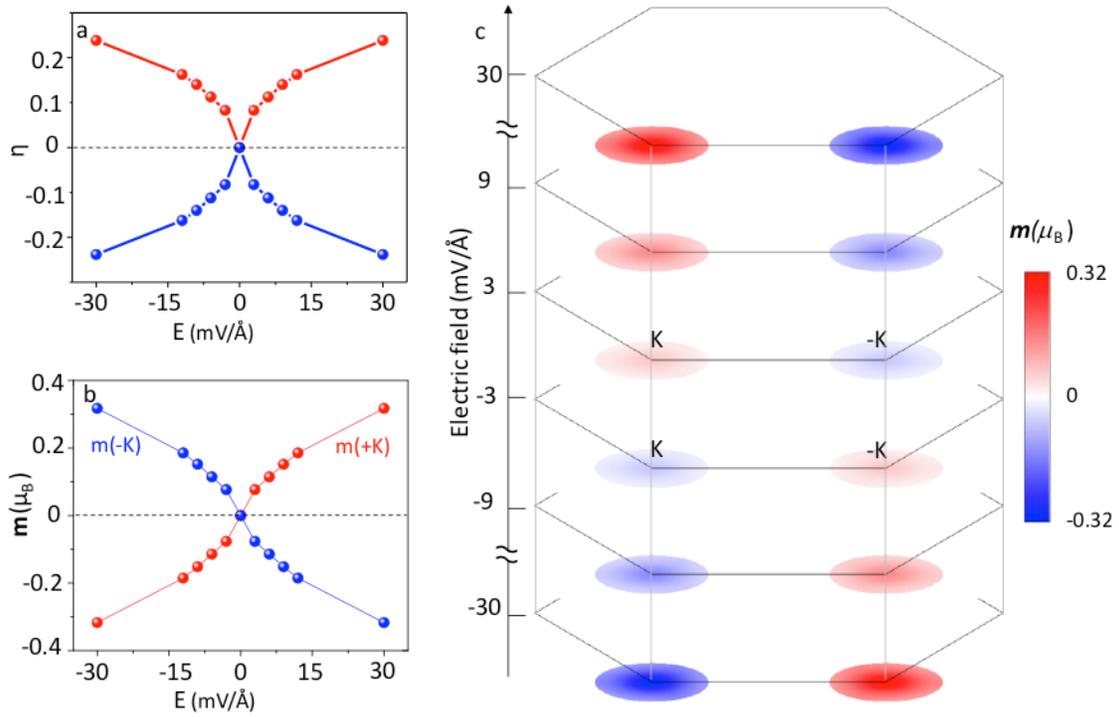

**Figure 4| DFT Calculation of magnetoelectric effect and associated PL polarization. a,** $\eta$ as a function of electric field. The positive (negative) value represents σ+ (σ-) excitation. **b**, **m** at $\pm K$ as a function of electric field, which shows that m is an odd function of electric field. **c,** color map of **m** as a function of electric fields near $\pm K$ points in the momentum

# Supplementary materials

# Electrical Tuning of Valley Magnetic Moment via Symmetry Control

Sanfeng Wu[1], Jason S Ross[2], Grant Aivazian[1], Aaron Jones[1], Zaiyao Fei[1], Gui-Bin Liu[3],

Wenguang Zhu[4-6], Di Xiao[4,7], Wang Yao[3], David Cobden[1], Xiaodong Xu[1,2†]

[†]Email: xuxd@uw.edu

**Table of Contents**:

**S1.** Polarization of bilayer devices at zero gate voltage

**S2.** Initial electron doping effect on bilayer devices

**S3.** Photoluminescence intensity as a function of gate voltage

**S4.** Ab initio Density Functional Theory Calculation

**S5**. Supplemental References

## S1. Polarization of bilayer devices at zero gate voltage

| Device # | B1 | B2 | B3 | B4 | B5 | B6 | B7 | B8 | B9 |
|---|---|---|---|---|---|---|---|---|---|
| η | 31% | 14% | 12% | 20% | 16% | 30% | 17% | 28% | 16% |

Table S1: We have measured a number of bilayer devices and all exhibit a nonzero degree of photoluminescence (PL) polarization (10% ~ 30%) at zero gate voltage. This observation indicates the existence of inversion symmetry breaking in bilayer $MoS_2$. The exact mechanism causing the inversion asymmetry is unclear but it is likely from the coupling to the substrates[1], consistent with the initial electron doping effect described in section **S2**.

## S2. Initial Electron Doping Effect on Bilayer Devices

In the main text, we show that PL polarization vanishes around $V_g$ = -60 V for device B2, where the top and bottom layers reach equal potential. Since the negative gate voltage depletes electrons or induces hole doping, the devices are initially electron doped as previous reported[2]. The electron doping is likely from trapped charges between $MoS_2$ and the $SiO_2$ substrate. All measured devices show the same trend for $\eta$ to decrease as $V_g$ decreases from positive to negative, consistent with initial electron doping; however, initial doping varies between devices. Figure S1 shows the degree of PL polarization from two devices under right-hand circularly polarized light excitation. Due to the heavy electron doping in these devices, the critical voltage for vanishing PL polarization is out of the accessible voltage range (beyond -100 V).

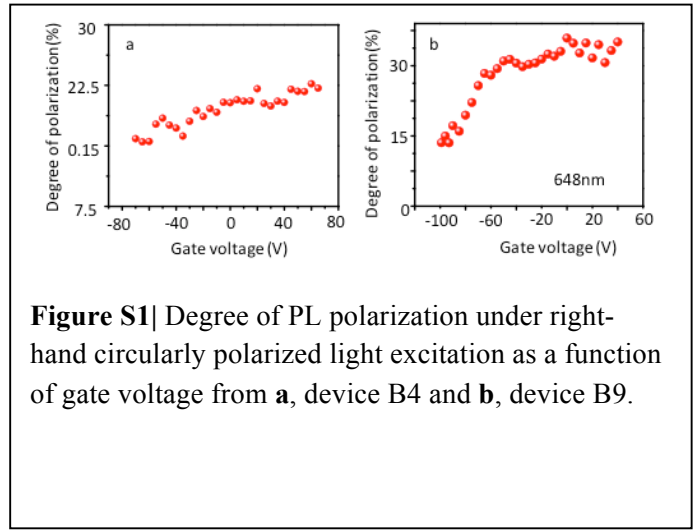

**Figure S1**| Degree of PL polarization under right-hand circularly polarized light excitation as a function of gate voltage from **a**, device B4 and **b**, device B9.

## S3. Photoluminescence as a Function of Gate Voltage

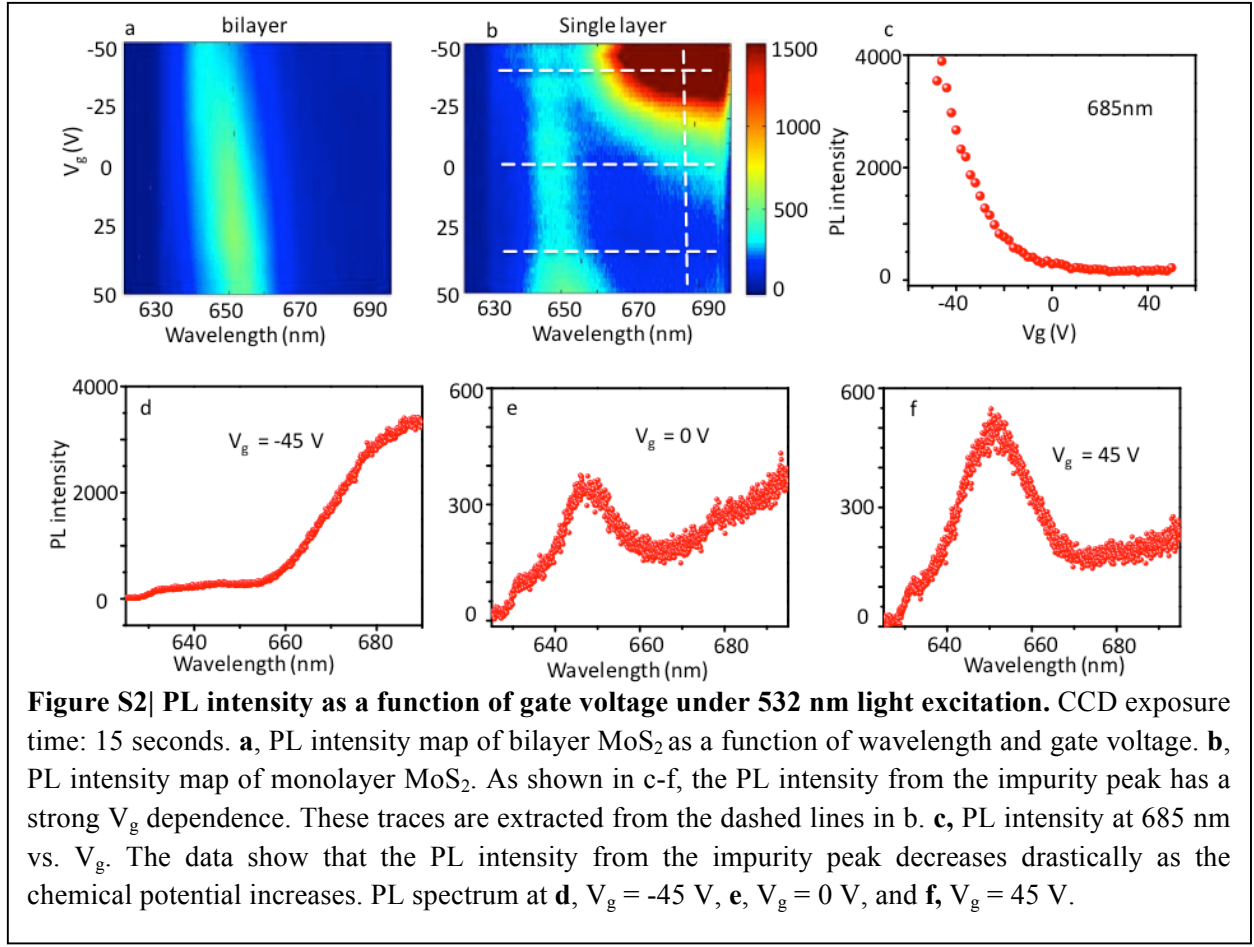

**Figure S2| PL intensity as a function of gate voltage under 532 nm light excitation.** CCD exposure time: 15 seconds. **a**, PL intensity map of bilayer $MoS_2$ as a function of wavelength and gate voltage. **b**, PL intensity map of monolayer $MoS_2$. As shown in c-f, the PL intensity from the impurity peak has a strong $V_g$ dependence. These traces are extracted from the dashed lines in b. **c,** PL intensity at 685 nm vs. $V_g$. The data show that the PL intensity from the impurity peak decreases drastically as the chemical potential increases. PL spectrum at **d**, $V_g$ = -45 V, **e**, $V_g$ = 0 V, and **f,** $V_g$ = 45 V.

## S4. Ab initio Density Functional Theory Calculation

Our first-principles density functional theory calculations were carried out using the projector augmented wave (PAW) method[3-4] with the local density approximation (LDA)[5] for exchange correlation as implemented in the Vienna Ab Initio Simulation Package (VASP)[6]. The Mo(4p,4d,5s) and S(3s,3p) electrons were treated as valence. A plane-wave energy cutoff of 600.0 eV was consistently used and a total of 80 bands were included in all the calculations. The supercell contains a 1×1 unit cell of $MoS_2$ bilayer and a vacuum region of 16 Å. A 36×36×1 special **k**-point mesh including the Γ point was used for integration over the Brillouin zone. Optimized atomic structures were achieved when forces on all the atoms were smaller than 0.001eV/Å. The optimized crystal parameter is 3.13 Å for $MoS_2$ bilayer. To generate a perpendicular external electric field, an artificial dipole layer was placed in the middle of the vacuum region [7].

## S5. Supplementary References